\documentclass[utf8]{FrontiersinHarvard} 

\usepackage{url,hyperref,lineno,microtype,subcaption}
\usepackage[onehalfspacing]{setspace}

\def\keyFont{\fontsize{8}{11}\helveticabold }
\def\firstAuthorLast{Chambers {et~al.}} 
\def\Authors{Glen Chambers\,$^{1,*}$, David B. Jess\,$^{1,2}$, Shahin Jafarzadeh\,$^{1}$, Michele Berretti\,$^{3,4}$, \\ Samuel D.{\,}T. Grant\,$^{1}$, Marco Stangalini\,$^{5}$, H. N. Smitha\,$^{6}$, \\ Damian J. Christian\,$^{2}$, Lu{\'{i}}s E.{\,}A. Vieira\,$^{7}$, Alisson Dal Lago\,$^{7}$, \\ and Fernando L. Guarnieri\,$^{7}$}

\def\Address{
$^{1}$Astrophysics Research Centre, School of Mathematics and Physics, Queen's University Belfast, Belfast BT7 1NN, Northern Ireland, UK \\
$^{2}$Department of Physics and Astronomy, California State University Northridge, Northridge, CA 91330, USA  \\
$^{3}$University of Trento, Via Calepina 14, 38122 Trento, Italy \\
$^{4}$University of Rome Tor Vergata, Department of Physics, Via della Ricerca Scientifica 3, 00133 Rome, Italy \\
$^{5}$ASI Italian Space Agency, Via del Politecnico snc, 00133 Rome, Italy \\
$^{6}$Max Planck Institute for Solar System Research, Justus-von-Liebig-Weg 3, 37077 G\"{o}ttingen, Germany \\
$^{7}$National Institute for Space Research (INPE/Brazil), S{\~{a}}o Jos{\'{e}} dos Campos, S{\~{a}}o Paulo 12227-010, Brazil
}
\def\corrAuthor{Glen Chambers}
\def\corrEmail{gchambers08@qub.ac.uk}

\usepackage{gensymb}

\begin{document}

\newcommand{\nadone}{Na{\,}{\sc{i}}{\,}D$_{1}$}
\newcommand{\nadtwo}{Na{\,}{\sc{i}}{\,}D$_{2}$}
\newcommand{\nadonetwo}{Na{\,}{\sc{i}}{\,}D$_{1}$/D$_{2}$}

\onecolumn
\firstpage{1}

\title[Integral field spectroscopy of a resonance cavity]{Standing oscillations in a resonant sunspot atmosphere captured by integral field spectroscopy} 

\author[\firstAuthorLast ]{\Authors} 
\address{} 
\correspondance{} 

\extraAuth{}

\maketitle

\begin{abstract}
\noindent 
\noindent The solar atmosphere is replete with magnetohydrodynamic wave activity, with magnetic structures such as sunspots and pores able to channel wave energy flux efficiently into the outer atmosphere. Steep density and temperature gradients at the base of the photosphere and top of the chromosphere provide ideal conditions for magnetoacoustic resonance cavities, enabling amplification of $\sim5$~mHz oscillatory power in sunspot atmospheres. Unfortunately, observational diagnosis of resonance cavities has largely been limited to spectral lines such as Ca~{\sc{ii}}~H/K and He~{\sc{i}}~10{\,}830~{\AA}, with no evidence yet from other chromospheric layers such as the {\nadonetwo} doublet. In this study we utilize the newly commissioned integral field unit, FRANCIS, to examine oscillatory characteristics spanning the formation heights of the {\nadonetwo} spectral lines and determine whether propagating and/or standing wave modes are present within a sunspot umbra. The RH1.5D radiative transfer code was used to estimate formation heights across three spectral windows corresponding to the {\nadone} line wing (line core $-300$~m{\AA}; $\approx355$~km), the {\nadone} line core ($\approx750$~km), and the {\nadtwo} line core ($\approx850$~km). Wavelet cross-correlation of line-core and line-bisector Doppler velocities enabled phase spectra to be generated as a function of atmospheric height, allowing the dominant $\sim5.5$~mHz oscillations to be classified as either propagating or standing-like. At the umbra–penumbra boundary we find propagating modes with energy fluxes of $\sim1.3 \times 10^{4}$~W{\,}m$^{-2}$ in the upper photosphere, decreasing to $\sim3.1 \times 10^{3}$~W{\,}m$^{-2}$ in the lower chromosphere, indicating a damping length of $L_d \approx 363$~km, comparable to the local density scale height. In contrast, near-zero velocity phase differences dominate regions of enhanced chromospheric power at the umbral center, providing observational evidence consistent with standing-wave behaviour and resonance-cavity dynamics in the lower chromosphere. These results demonstrate the suitability of solar integral field units for spatially mapping sunspot wave properties, with the {\nadonetwo} lines providing a novel vantage point for resonance cavity and energy flux diagnostics in the lower solar atmosphere.

\section{}

\tiny
 \keyFont{ \section{Keywords:} Magnetohydrodynamics (MHD), Solar chromosphere, Solar instruments, Solar magnetic fields, Solar oscillations, Solar photosphere, Sunspots} 
\end{abstract}

\section{Introduction}
Since as early as the 1960s, oscillatory phenomena have been observed in the solar atmosphere \citep{1960IAUS...12..321L}, with the study of Doppler velocity and intensity time series from spectral lines spanning the bulk of the electromagnetic spectrum confirming the ubiquity of these oscillations throughout the lower solar atmosphere \citep{1970ApJ...162..993U, 1975A&A....44..371D, 1978SoPh...56....5R, 1982SoPh...75....3S, 1992A&A...257..287T, 2003ApJ...599..626B, 2012ApJ...746..183J, 2015SSRv..190..103J, 2023LRSP...20....1J}. The photosphere displays global pressure modes ($p$-modes), which are omnipresent waves that exist between frequencies of $1-5$~mHz and are effectively acoustic in nature in the absence of magnetic fields \citep{1962ApJ...135..474L, 1978A&A....70..345C, 1997ApJ...488..462R}. However, $p$-mode oscillations are also able to be effectively guided in the presence of magnetic field configurations in the solar atmosphere, e.g., magnetic bright points \citep{1995ApJ...454..531B, 2009Sci...323.1582J, 2010A&A...511A..39U, 2013A&A...549A.116J, 2017ApJS..229...10J, 2024A&A...690A.363B, 2025A&A...695L..11S}, solar pores \citep{2011ApJ...729L..18M, 2018ApJ...857...28K, 2021RSPTA.37900172G, 2022ApJ...938..143G, 2024A&A...688A...2J, 2024ApJ...975...45S, 2026ApJ...997..197B}, and sunspots \citep{1997SoPh..172...69H, 2007ApJ...671.1005B, 2007A&A...473..943J, 2021A&A...649A.169S, 2022ApJ...924..100C, 2022ApJ...938..154M, 2023MNRAS.525.4815R}. Here, the embedded oscillations take on properties associated with magnetohydrodynamic (MHD) waves, with compressive modes essentially becoming magnetoacoustic \citep{1983SoPh...88..179E}.   

Within magnetic structures with longer lifetimes, such as sunspots and pores, frequencies of $\sim$3~mHz are observed to dominate at photospheric heights \citep{1970ApJ...162..993U}, while within the chromosphere the dominant frequency shifts to $\sim$5~mHz \citep{1991A&A...250..235F, 2006ApJ...640.1153C, 2010ApJ...722..131F, 2020A&A...640A...4F}. Of course, this behavior is not strictly universal, since previous studies have uncovered dominant frequencies of $\sim$5~mHz at photospheric heights \citep{2022NatCo..13..479S, 2025A&A...697A.156B}, highlighting dependencies on both the oscillatory driver and the atmospheric structures acting as a wave conduit. The transition in dominant periodicity from $\sim$5~minutes ($\sim$3~mHz) to $\sim$3~minutes ($\sim$5~mHz) between the photosphere and chromosphere is often attributed to the manifestation of an acoustic cutoff frequency at $\sim$5.3~mHz \citep{1984ARA&A..22..593D, 1991ApJ...373..308D, 1991A&A...250..235F, 1992A&A...266..532F, 1998MNRAS.298..464V}. For waves that propagate between the photosphere and chromosphere, energy fluxes of the oscillations at different atmospheric heights have been studied and linked to effective damping lengths. For example, \citet{2021RSPTA.37900172G} and \citet{2017ApJ...847....5K} examined waves in solar pores and a sunspot, respectively, and found energy fluxes of $10^3 - 10^5$~W{\,}m$^{-2}$ in the lower solar atmosphere, with damping lengths on the order of $\sim$270~km, which is of a similar size to the density scale height in the photosphere and showcases the rapid damping experienced by these propagating MHD waves. 

However, in recent years, evidence continues to grow in support of a resonance cavity being a viable alternative mechanism to help amplify frequencies on the order of 5~mHz within the chromosphere \citep{2020NatAs...4..220J, 2021NatAs...5....5J, 2020ApJ...900L..29F, 2024MNRAS.529..967S, 2025ApJ...986..180S}. Here, structures that demonstrate steep temperature/density gradients at both the base of the photosphere and the top of the chromosphere provide strong reflective boundaries that amplify specific frequencies trapped within the resonance cavity \citep{1979SoPh...62..227H, 2011ApJ...728...84B, 2015A&A...580A.107S}. As a resonance cavity creates trapped, i.e., {\textit{standing}} oscillations, phase analyses of the induced oscillation patterns between neighboring atmospheric heights are able to ascertain whether the detected oscillations are truly standing \citep{2002ApJ...564..508R, 2014A&A...563A..12D, 2021A&A...645L..12F, 2025A&A...693A.165F, 2022ApJ...938..143G}, hence providing additional evidence of the existence of resonance cavities in the lower solar atmosphere. 

Towards the boundary of sunspot umbrae, the magnetic field lines begin to become more heavily inclined into the penumbral regions, reaching values of $\sim$90$^{\circ}$ (i.e., parallel to the solar surface) within the chromospheric layers due to the rapid expansion and curvature of the field lines \citep{2016AN....337.1040L, 2018NatPh..14..480G}. Within such regions with non-vertical magnetic field lines, the acoustic cutoff frequency becomes modulated by the effective gravity \citep{2006ApJ...647L..77M, 2012ApJ...746..119R, 2012ApJ...756...35R, 2021ApJ...923..225M}, hence progressively enabling lower frequencies to propagate more effectively with increasing field inclinations \citep{2013ApJ...779..168J, 2016AN....337.1040L}. This enables additional wave phenomena, such as running penumbral waves \citep[RPWs;][]{1972SoPh...27...71G}, to become visible in chromospheric intensity and Doppler time series \citep{1972ApJ...178L..85Z, 1997ApJ...478..814B, 2000SoPh..196..129K, 2004A&A...424..671K}. 

Traditionally, chromospheric diagnostics are often derived from hydrogen (e.g., H$\alpha$ and H$\beta$), magnesium (e.g., Mg{\,}{\sc{ii}}{\,}h/k and Mg{\,}{\sc{I}}{\,}b$_{2}$) and ionized calcium (e.g., Ca{\,}{\sc{ii}}{\,}H/K and Ca{\,}{\sc{ii}}{\,}8542{\,}{\AA}) spectral observations. However, two additional spectral lines of interest are the {\nadone} (589.592~nm) and {\nadtwo} (588.995~nm) Fraunhofer lines that form in the solar spectrum as a sodium doublet. The {\nadone} and {\nadtwo} absorption lines originate from the $3{\mathrm{S}}_{1/2} \rightarrow 3{\mathrm{P}}_{1/2}$ and $3{\mathrm{S}}_{1/2} \rightarrow 3{\mathrm{P}}_{3/2}$ electronic transitions, respectively \citep{1992A&A...265..237B}. The difference in $J$-level transition leads to divergence in how their characteristics are affected by polarization, i.e., only the {\nadtwo} transition has $J \geq 1$ \citep{2016A&A...591A..60B}. These differences also extend to the formation heights associated with the two spectral lines. Initial estimations of the {\nadone} and {\nadtwo} formation heights proved challenging, with early observational estimates suggesting formation heights between $1300-1700$~km \citep{1976SvA....20..201A}. More recent estimations using the radiative modeling of response functions suggest formation heights slightly below 1000~km, ranging from 300~km to 850~km \citep{2001A&A...371.1128E, 2004ApJ...613L.185F, 2010ApJ...709.1362L}. Nevertheless, formation heights synonymous with the upper photosphere and lower chromosphere open up new diagnostic potential when examining the coupling of oscillatory behavior in the lower solar atmosphere. 

Various studies in the 1980s utilized the {\nadonetwo} lines to demonstrate the presence of wave motions in the vicinity of sunspots by investigating intensity and Doppler velocity perturbations \citep{1981A&A...102..147K}. Variations in the phase angles between the {\nadonetwo} signals with frequency were taken as evidence of upward wave propagation through the sunspot atmosphere \citep{1983A&A...123..263V}, with \citet{1990Ap&SS.170...43M} noting distinct differences between the wave signatures in umbral and penumbral environments. The 3-minute oscillations typically found at chromospheric heights in sunspot atmospheres were also confirmed to exist within the sodium doublet \citep{1991A&A...241..212M}. Since then, the {\nadonetwo} spectral lines have been relatively underutilized in the analyses of sunspot oscillations, although numerous studies have exploited these lines to examine other dynamic features, such as resonant scattering in magnetic bright points \citep{2002ESASP.477..147M, 2010ApJ...719L.134J} and the pressure balance and flow gradients present during solar flares \citep{2016ApJ...832..147K}. Importantly, these studies note the need for longer duration observations with new instrumentation that provides higher temporal, spatial, and spectral resolutions in order to unequivocally investigate the dynamic spectral line shapes associated with the {\nadonetwo} doublet in the lower solar atmosphere. 

With the need for not only high-resolution spatial imaging, but also spectral and polarimetric capabilities too, various instrumentation solutions have been proposed and employed, including those on ground-based telescopes such as the Swedish Solar Telescope \citep[SST;][]{2003SPIE.4853..341S, 2017psio.confE..85S}, the Dunn Solar Telescope \citep[DST;][]{1969S&T....38..368D} and the GREGOR solar telescope \citep{2012AN....333..796S}, as well as onboard balloon facilities such as SUNRISE \citep{Sunrise_2010ApJ...723L.127S} and space-borne missions such as Solar Orbiter \citep{oribter_2020A&A...642A...1M}. More long-established spectropolarimeter solutions have included Fabry-P{\'{e}}rot interferometers \citep{2006SoPh..236..415C, 2008ApJ...689L..69S, Fab_P_2023Ap&SS.368...55B} and scanning slit-based spectrographs \citep{2008AGUSMSH31A..11J, 2012SPIE.8446E..6XD}. While these instruments successfully allow the observer to capture both the spatial and spectral domains, they are unable to do so simultaneously, creating challenges when studying features with evolutionary time frames shorter than typical scanning or raster periods. However, more recent developments in solar instrumentation, reviewed by \citet{iggyfell..58h2417I}, have led to the implementation of integral field solutions, which solve the simultaneity issue by placing all three dimensions ([$x, y, \lambda$]) onto a single two-dimensional imaging detector, albeit with a necessary reduction in the observable field of view. Options for integral field units (IFUs) include microlens arrays, image slicers, and fibre-bundle reformatting. While the adoption of IFUs by the solar physics community is relatively recent, several are now operational at major solar telescopes, including the Diffraction-Limited Near-Infrared Spectropolarimeter \citep[DL-NIRSP;][]{dlnirsp} at the Daniel K. Inouye Solar Telescope \citep[DKIST;][]{2020SoPh..295..172R}, the GREGOR Infrared Spectrograph \citep[GRIS;][]{gris} at the GREGOR Solar Telescope, the Microlensed Hyperspectral Imager \citep[MiHI;][]{2022A&A...668A.149V}, and the Fibre Resolved OpticAl and Near-Ultraviolet Czerny-Turner Imaging Spectropolarimeter \citep[FRANCIS;][]{2023SoPh..298..146J} at the DST. 

In this paper, we utilize cutting-edge integral field spectroscopy of a sunspot captured by the FRANCIS instrument within the $580.758-597.250$~nm wavelength range to investigate the oscillation characteristics embedded within, {\textit{and between}}, the {\nadone} and {\nadtwo} line pair. In particular, we will investigate the oscillatory phase relationships between line-core and bisector Doppler velocities to determine whether the waves are standing or propagating within the sunspot atmosphere, which will enable us to shed light on whether {\nadonetwo} doublet spectroscopy offers observers a unique vantage point when establishing the presence of resonance cavities in the lower solar atmosphere. 

\section{Observations}

\begin{figure}[t!]
	\includegraphics[width=1\textwidth]{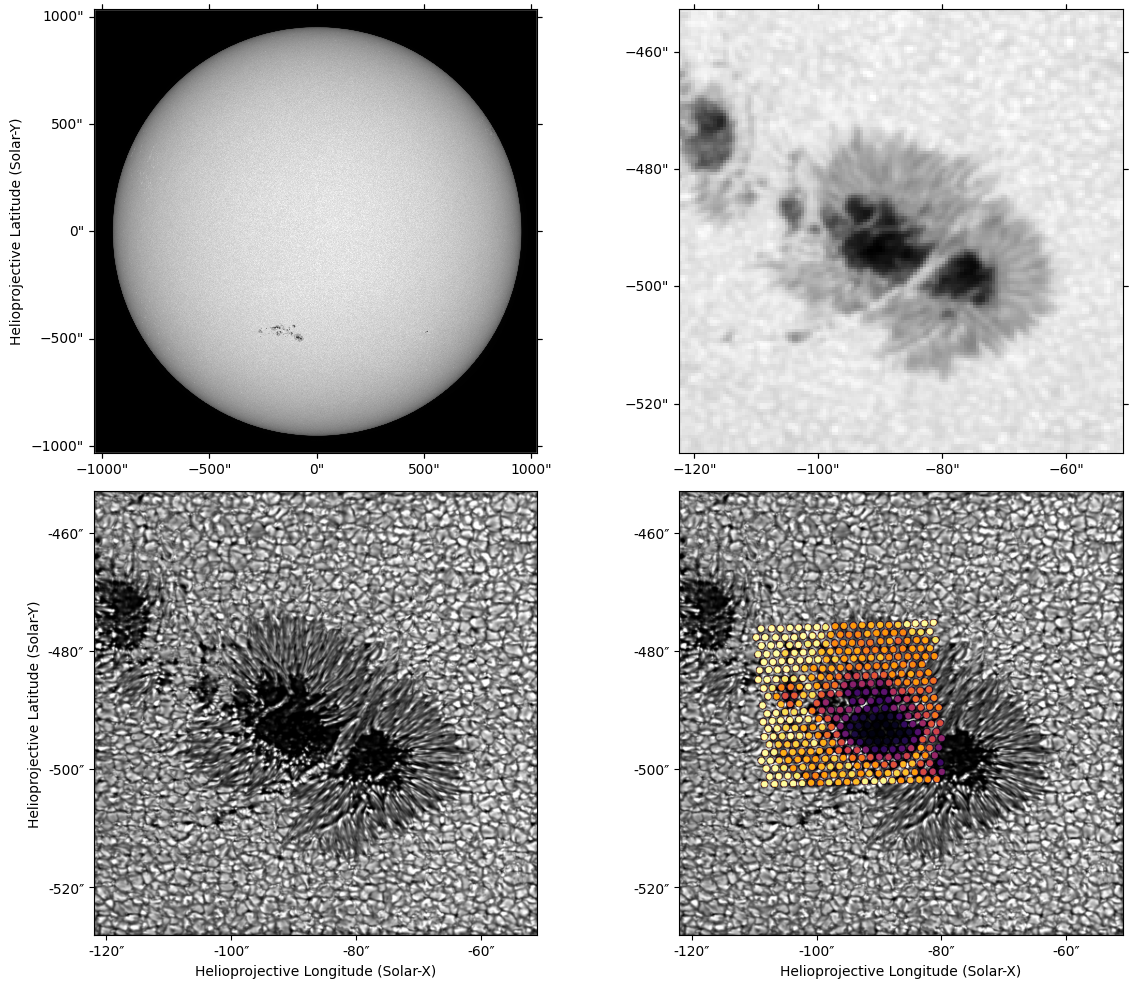}
	\centering
	\caption{A full disk SDO/HMI continuum image acquired at 17:12~UT (upper left). The upper-right panel displays a sub-field of the SDO/HMI continuum image that is identical to that of the ROSA time series. The lower-left panel is a ROSA continuum image, also obtained at 17:12~UT, while the lower-right panel shows the same ROSA continuum image but now overplotted with the two-dimensional FRANCIS fiber array, where the color of each fiber (using a black-red-yellow-white color scale) reflects the continuum intensity extracted from between the {\nadone} and {\nadtwo} spectral lines.}
	\label{fig:FoV}
\end{figure}

The primary observations utilized in this study were captured using a series of instruments at the DST in the Sacramento Peak Mountains, New Mexico, USA in late August 2022. The main science instrument used was the FRANCIS IFU, with context images provided by the Rapid Oscillations in the Solar Atmosphere \citep[ROSA;][]{2010SoPh..261..363J} and Hydrogen-alpha Rapid Dynamics camera \citep[HARDcam;][]{2012ApJ...757..160J} imaging systems. This is the same dataset used for science verification of the FRANCIS instrument \citep{2023SoPh..298..146J}, with FRANCIS observations spanning the $580.758-597.250$~nm wavelength range, with a spectral sampling of $\Delta \lambda = 0.008$~nm, and were captured between 18:03$-$18:39~UT on 2022 August 29. Active region NOAA~AR~13089, located at heliocentric coordinates ($-86''$, $-491''$), or S23.9E5.7 in Stonyhurst heliographic coordinates, was observed under excellent seeing conditions, with high-order adaptive optics \citep{2004SPIE.5490...34R} further improving the resulting image quality.

Over the course of the observations, a series of 50{\,}000 Stokes~$I$ spectral images, with a frame rate of 23~s$^{-1}$, were obtained. The $20\times20$ two-dimensional fiber bundle of the FRANCIS instrument covered an approximate $30\times30$~arcsec$^{2}$ field-of-view, resulting in $\approx1{\,}.{\!\!}''5$ across the diameter of each fiber, providing the best compromise between spatial sampling and overall field-of-view coverage that enabled a large portion of NOAA~AR~13089 to be captured. The FRANCIS spectral images were reduced and reformatted back into four-dimensional cubes ([$x, y, \lambda, t$]) following the methods described in \citet{2023SoPh..298..146J}, with spectral intensities normalized by the continuum level, $I_{c}$. However, an additional correction to the imaged spectral curvature was performed using an established `smile detection' algorithm provided by the Spectroflat package \citep{2024A&A...687A..22H}. Finally, as this study is primarily interested in oscillations within the frequency range of $\sim1-100$~mHz (i.e., far exceeding the upper range of the typical $p$-mode spectrum), spectra were temporally binned (47 $\rightarrow$ 1) to further improve the signal-to-noise, resulting in a new cadence of 1.98~s, corresponding to a Nyquist frequency of approximately 250~mHz. 

Following the initial wavelength calibration procedure outlined in \citet{2023SoPh..298..146J}, we applied an additional calibration step before removing the instrumental asymmetric broadening described in Section~{\ref{sec:timeseriesanalysis}}. Here, a time-averaged spectrum was constructed from the full observational sequence, and the minima of $\sim$15 absorption features distributed across the observed spectral range were identified. The corresponding features were located in the FTS solar atlas \citep{1978fsoo.conf...33B}, and a revised wavelength grid was derived by mapping the observed line minima onto their atlas counterparts, such that the time-averaged spectrum and the FTS atlas are brought into agreement across the full spectral range. Since this procedure is applied to the temporally averaged profile, in which any residual Doppler shifts average to near zero over the duration of the observations, the resulting wavelength solution is representative of the rest wavelength scale. The wavelength correction determined from the averaged spectrum was subsequently applied uniformly to each individual time step, ensuring a consistent and stable wavelength calibration across the full data set.

ROSA context images were provided via a 5.2~nm bandpass continuum filter centered at 417.0~nm, which were acquired with a cadence of 33~ms, providing a speckle reconstructed final cadence of 2.11~s \citep[64 $\rightarrow$ 1 restorations;][]{2008A&A...488..375W}. The ROSA continuum images were taken with a platescale of $0{\,}.{\!\!}''077$ per pixel, providing a field-of-view size equal to $77''\times77''$. Additional supporting images from the Atmospheric Imaging Assembly \citep[AIA;][]{2012SoPh..275...17L} and the Helioseismic and Magnetic Imager \citep[HMI;][]{2012SoPh..275..207S} on-board the Solar Dynamics Observatory \citep[SDO;][]{2012SoPh..275....3P} are used to provide both full-disk context images for co-alignment and photospheric vector magnetograms for the computation of extrapolated magnetic fields from the observed active region. Co-alignment between the FRANCIS two-dimensional fiber array and the contextual SDO/HMI and ROSA images was performed using the back-reflected image from the fiber ferrule. The full disk SDO/HMI continuum intensity image can be seen in the upper-left panel of Figure~{\ref{fig:FoV}}, then cropped to the relevant sub-field in the upper-right panel. The corresponding ROSA continuum image is displayed in the lower-left panel of Figure~{\ref{fig:FoV}}, while the lower-right panel displays the same ROSA context image but with the co-aligned two-dimensional FRANCIS fiber array overplotted. Here, the intensities of each FRANCIS fiber are extracted from the continuum that bridges the {\nadone} and {\nadtwo} spectral lines. 

\begin{figure}[t!]
	\includegraphics[width=1\textwidth]{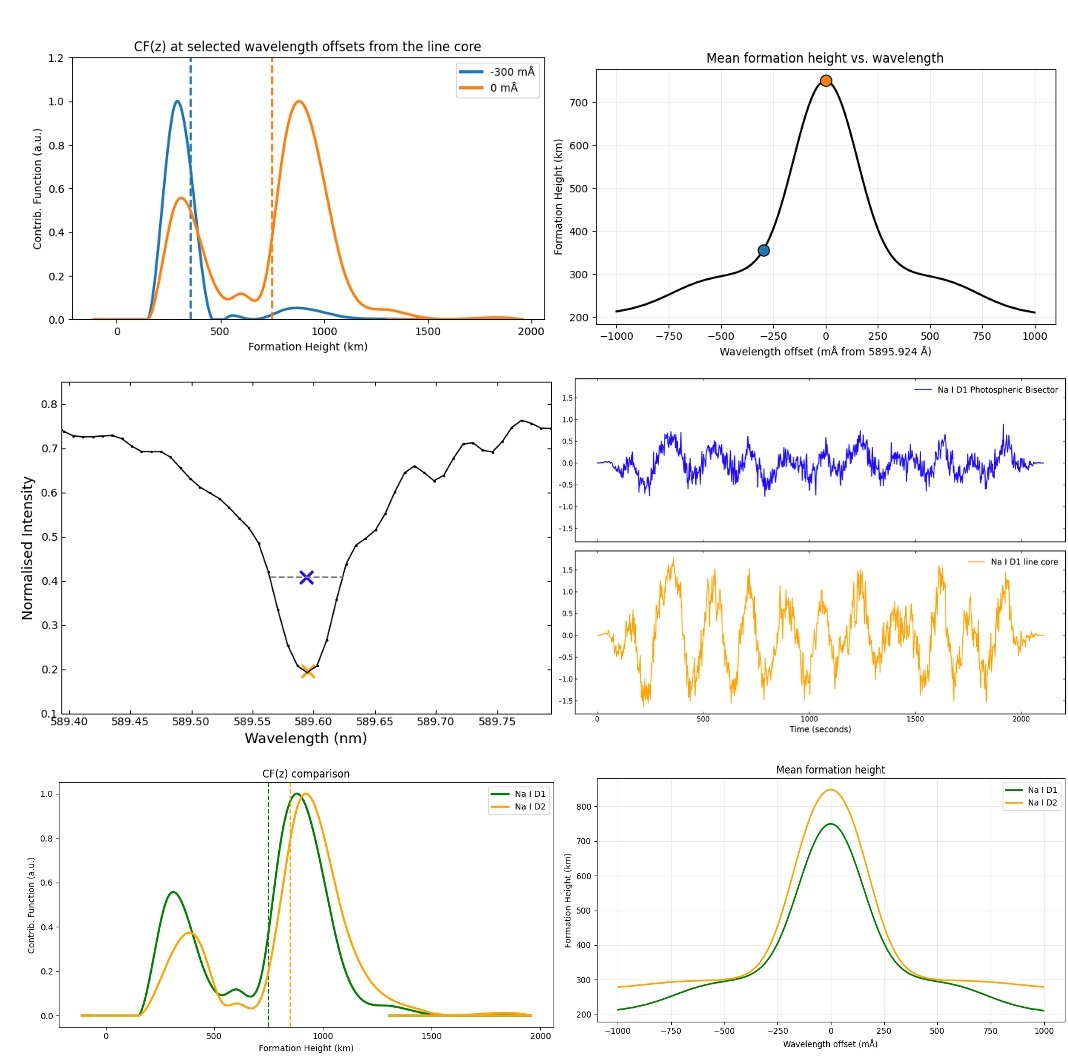
    }
	\centering
	\caption{The upper-left panel displays the contribution functions for the {\nadone} line core (solid orange line) and the {\nadone} wing ($-300$~m{\AA}; solid blue line), where the mean formation heights for each have been highlighted using a vertical dashed line. The upper-right panel displays the estimated formation heights (in km) across the {\nadone} spectral line, where the orange and blue circles indicate the wavelengths of interest used in our present study (i.e., {\nadone} line core in orange, with the {\nadone} $-300$~m{\AA} in blue), which correspond to the contribution functions depicted in the upper-left panel. The middle-left panel shows a typical umbral {\nadone} Stokes~$I/I_{c}$ spectrum (i.e., normalized to the quiet Sun continuum intensity, $I_{c}$), where the line core and 75\% line depth bisector wavelengths are indicated using orange and blue crosses, respectively. Velocity time series, corresponding to line-core Doppler velocities (orange line) and 75\% line depth bisector velocities (blue line), are shown in the middle-right panel as a function of time across the observing sequence.}  
	\label{fig:contribution}
\end{figure}

\section{Analysis and Discussion}
\subsection{Spectral Line Modeling}
In order to better estimate the formation heights of the two {\nadonetwo} absorption lines, particularly within the atmosphere of a highly magnetic sunspot, we first compute the contribution functions of each Na{\,}{\sc{i}} line using the RH1.5D radiative transfer code \citep{2001ApJ...557..389U, 2015A&A...574A...3P}. The RH1.5D code computes solutions for solar atmospheric radiative transfer equations alongside statistical equilibrium equations \citep{2017ApJS..229...10J}. The Maltby `M' model atmosphere \citep{1986ApJ...306..284M} is used as this best represents the sunspot umbral environment we are examining in our present work. Next, the contribution functions and spectral profiles are convolved with a Gaussian matching the spectral instrumental profile of FRANCIS. 

The upper-left panel of Figure~\ref{fig:contribution} shows two contribution functions for the {\nadone} line, corresponding to the line core (orange line) and the line core $-300$~m{\AA} (blue line). We use these contribution functions because they directly show how each atmospheric layer contributes to the emergent intensity, providing an intuitive and computationally efficient estimate of representative formation heights. A wavelength separation of $300$~m{\AA} was chosen to ensure the resulting formation height separation was sufficiently large, while also avoiding unnecessarily large wavelength separations that result in contamination from spectral blends far into the {\nadonetwo} line wings (see, e.g., the lower-left panel of Figure~{\ref{fig:contribution}}). The centroids of the contribution functions in geometric height can be taken as an estimation of the formation height of the line, providing formation heights of $\sim355 \pm 40$~km and $\sim750 \pm 50$~km for the {\nadone} line wing and line core, respectively. The uncertainty in the estimated formation heights of each line position was calculated using the spectral resolution of FRANCIS (derived from the spectral sampling of $\Delta \lambda = 0.008$~nm), which is subsequently propagated through the RH1.5D code to provide the degree of uncertainty in the estimated formation heights. Incorporating two wavelength positions for the {\nadone} spectral line (line core and line core $-300$~m{\AA}), alongside the formation height of the {\nadtwo} line core, we are provided with three distinct formation heights from which to extract and analyze wave behavior. These formation heights correspond to $\sim 355 \pm 40$~km ({\nadone} line core $-300$~m{\AA}), $\sim 750 \pm 50$~km ({\nadone} line core), and $\sim 850 \pm 50$~km ({\nadtwo} line core). However, we note that the formation heights derived here are independent of the Wilson depression \citep{1774RSPT...64....1W}, whereby the height of the visible umbral photosphere is often depressed by $\sim500$~km when compared to the surrounding quiet Sun \citep{1993A&A...277..639S, 2018A&A...619A..42L}. Hence, the heights employed throughout this work are representative of the formation heights above the umbral photosphere and are not necessarily equivalent to those found in quiet Sun locations. 

We emphasize that these formation heights are intended as approximate, model-dependent reference values. In the present study, contribution functions are used because they directly quantify how different atmospheric layers contribute to the emergent intensity, thereby providing a physically motivated and computationally efficient estimate of representative line-core formation heights. By contrast, response functions describe the sensitivity of the emergent signal to perturbations in specific atmospheric parameters, and are more naturally suited to detailed sensitivity or inversion analyses. Since our aim here is to estimate representative formation heights rather than to characterize the response to individual perturbations, the use of contribution functions is appropriate for the present work. We further note that, because the RH1.5D calculations are based on the Maltby `M' umbral atmosphere, the resulting geometric heights are most representative of the umbral core and should be regarded as first-order approximations when applied to more inclined or thermodynamically distinct regions, such as the umbra--penumbra boundary. We also note that the {\nadonetwo} line profiles are formed in a strongly magnetized umbral atmosphere, hence the RH1.5D synthesis used here includes Zeeman splitting in the emergent Stokes~$I$ profiles as this may influence the detailed line shape and hence the exact wavelength or intensity level associated with a given bisector position. This effect is particularly relevant for the 75\% line-depth bisector diagnostic. As a result, the bisector-derived formation height should be treated as an approximate reference value rather than a sharply defined geometrical layer. This uncertainty propagates into quantities that depend directly on height separation, such as phase speeds, energy fluxes, and damping lengths, but it does not alter the directly measured phase relationships. 

\subsection{Magnetic field extrapolations}
When observing multiple spectral lines, and therefore comparing wave activity across multiple atmospheric heights, it is essential to understand how the magnetic field lines vary between the observed layers. If there is significant inclination of the magnetic field between atmospheric layers, then the MHD waves observed in one spatial location in the photosphere may no longer remain co-spatial when viewed from the chromosphere. Therefore, it was deemed necessary to perform magnetic field extrapolations from the SDO/HMI vector magnetograms in order to determine how the umbral waveguides connect across the sunspot atmosphere and whether photospheric/chromospheric dynamics are indeed co-spatial and remain captured by the same FRANCIS fiber. 

\begin{figure}[t!]
	\includegraphics[width=0.9\textwidth]{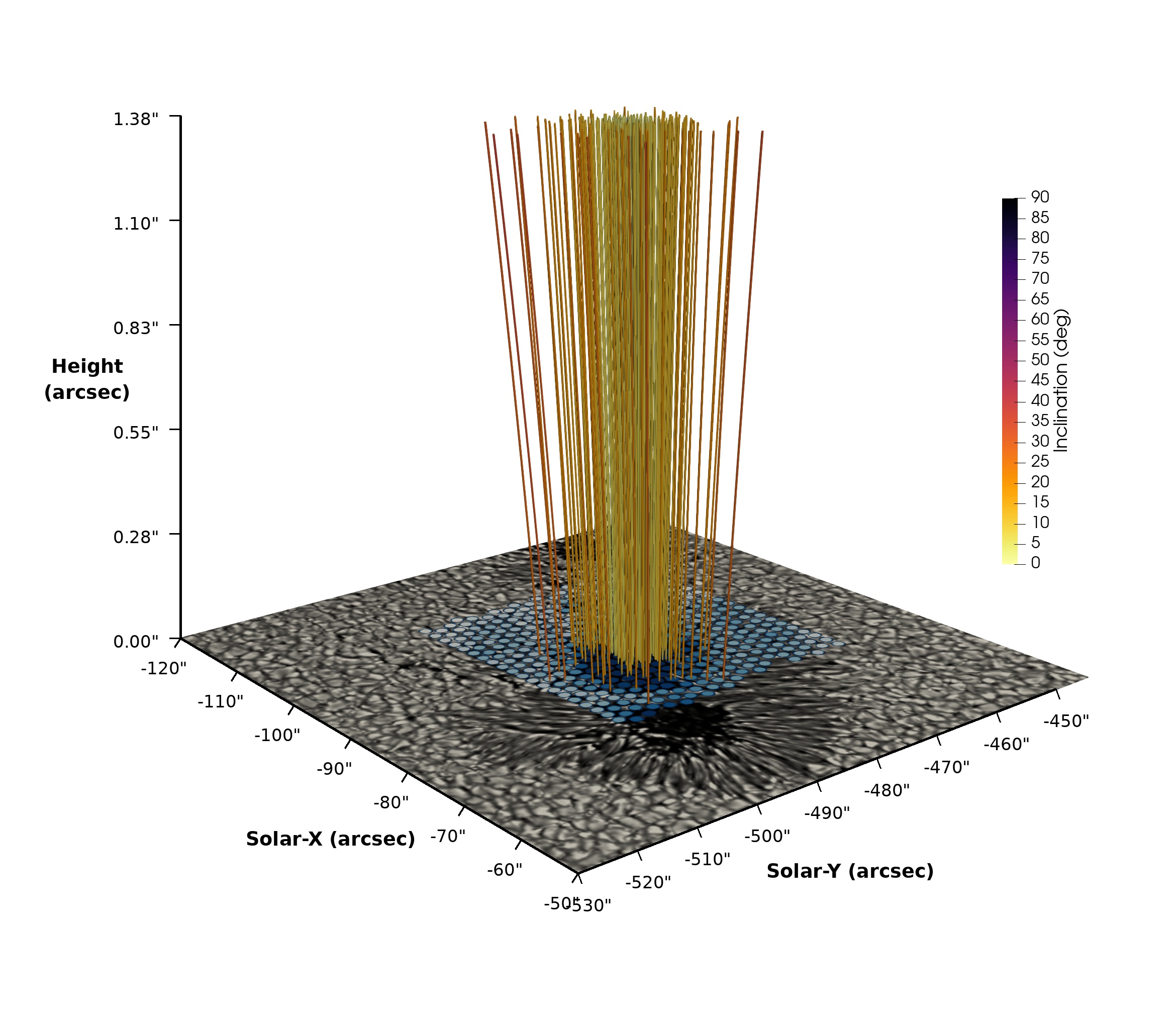}
	\centering
	\caption{Three-dimensional visualization of extrapolated magnetic field lines protruding from the umbral region within the photosphere that is co-spatial with the FRANCIS fiber array, extending to 1000~km (i.e., into the lower chromosphere). The base layer is a context ROSA continuum intensity image showing the observed sunspot using a black/white color table, while the FRANCIS fiber array is overplotted using a blue/white color table, where the intensity associated with each fiber position is extracted from a continuum wavelength between the {\nadone} and {\nadtwo} spectral lines. Extrapolated magnetic field lines from the VCA-NLFFF code are overplotted, where the inclination angles (in degrees) from the solar normal are displayed using a yellow-red-indigo color scale. The dominant color of the extrapolated field lines appears yellow-orange due to the low inclination angles associated with umbral core locations.}
	\label{fig:extrapolation}
\end{figure}

\begin{figure}[t!]
	\includegraphics[width=1\textwidth]{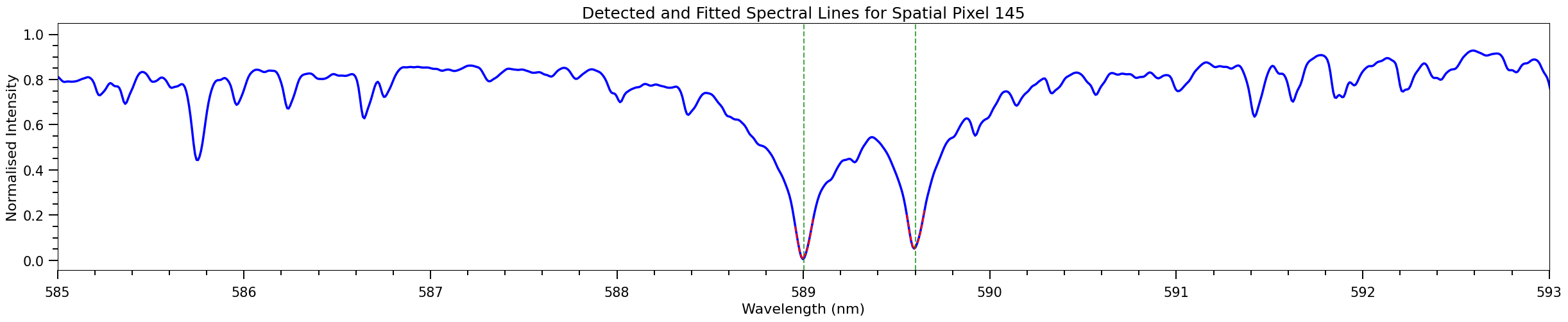}
	\centering
	\caption{A sample umbral spectrum from the FRANCIS instrument, centered on the {\nadonetwo} absorption line doublet and spanning a wavelength range of $585-593$~nm, is shown using a solid blue line. The overplotted red lines represent the fits of asymmetric Voigt profiles applied to each of the {\nadonetwo} spectral lines, while the vertical green dotted lines indicate the centroids of each of the fitted profiles.}
	\label{fig:spectra}
\end{figure}

To perform magnetic field extrapolations of the SDO/HMI vector magnetograms, the Vertical-Current Approximation Nonlinear Force-Free Field \citep[VCA-NLFFF;][]{2016ApJS..224...25A, 2016ApJ...826...61A} extrapolation code was utilized. This VCA-NLFFF code employs an SDO/HMI full disk vector magnetogram and decomposes it into its magnetic source components. Accompanying EUV/UV images from SDO/AIA are also taken as inputs to trace loops at coronal heights, which outline the magnetic field geometries at atmospheric heights exceeding several thousand km. The model then uses both sets of images to calculate a nonlinear force-free field solution, returning vector magnetic field information as a function of atmospheric height that enables inclination angles to be estimated from the expected formation heights of the {\nadonetwo} FRANCIS observations. 

To better visualize the outputs from the VCA-NLFFF magnetic field extrapolations, it is possible to trace individual field lines as they permeate through the solar atmosphere. Figure~{\ref{fig:extrapolation}} shows the magnetic structure relevant to the sunspot under investigation, where only magnetic field lines anchored into the umbral regions sampled by the two-dimensional FRANCIS fiber bundle are depicted. The cutout shown in Figure~{\ref{fig:extrapolation}} covers atmospheric heights spanning $0-1000$~km, which is consistent with the expected range of formation heights stemming from the {\nadonetwo} contribution functions. The inclination angles of the magnetic field, particularly towards the center of the sunspot umbra, are typically $<10^{\circ}$, highlighting the near-vertical nature of the magnetic field lines in the lower umbral atmosphere. 

When mapped onto the solar disk, a FRANCIS fiber core has a spatial diameter of $1{\,}.{\!\!}''1$ ($1{\,}.{\!\!}''5$ when the cladding is considered). Given the maximum height separation between the {\nadtwo} line core ($850 \pm 50$~km) and the {\nadone} line core $-300$~m{\AA} ($355 \pm 40$~km) is on the order of $500$~km, a simple trigonometric calculation reveals that inclination angles of $<38.7^{\circ}$ are sufficient to prevent the magnetic field lines from being displaced by more than the radius of a FRANCIS fiber core (i.e., $<0.55$~arcseconds) as they move upwards through the sunspot atmosphere (i.e., preventing fiber-to-fiber contamination of signals spanning the range of {\nadonetwo} formation heights). From examination of the VCA-NLFFF extrapolation outputs, such extreme angles are not identified at the heights corresponding to the {\nadonetwo} mean formation heights. Therefore, when comparing observations from both {\nadonetwo} lines, it can be reasonably assumed that the source features remain isolated to the same FRANCIS fiber across the sampled height range. However, the observed sunspot is situated at $\mu \approx 0.85$, so the observer's line of sight is inclined by $\sim31^{\circ}$ with respect to the local solar normal. Since the extrapolated umbral magnetic field is close to vertical in the umbral core, with inclinations typically below $\sim10^{\circ}$ over the relevant height range, the measured LOS velocities should be regarded as projected components of the field-aligned velocity. Consequently, the quoted velocity amplitudes, and hence the derived energy fluxes, are lower-limit estimates. This projection does not materially affect the phase-based distinction between standing-like behavior in the umbral core and propagating behavior near the umbra-penumbra boundary, but it should be considered when interpreting absolute flux values.

\begin{figure}[t!]
	\includegraphics[width=1\textwidth]{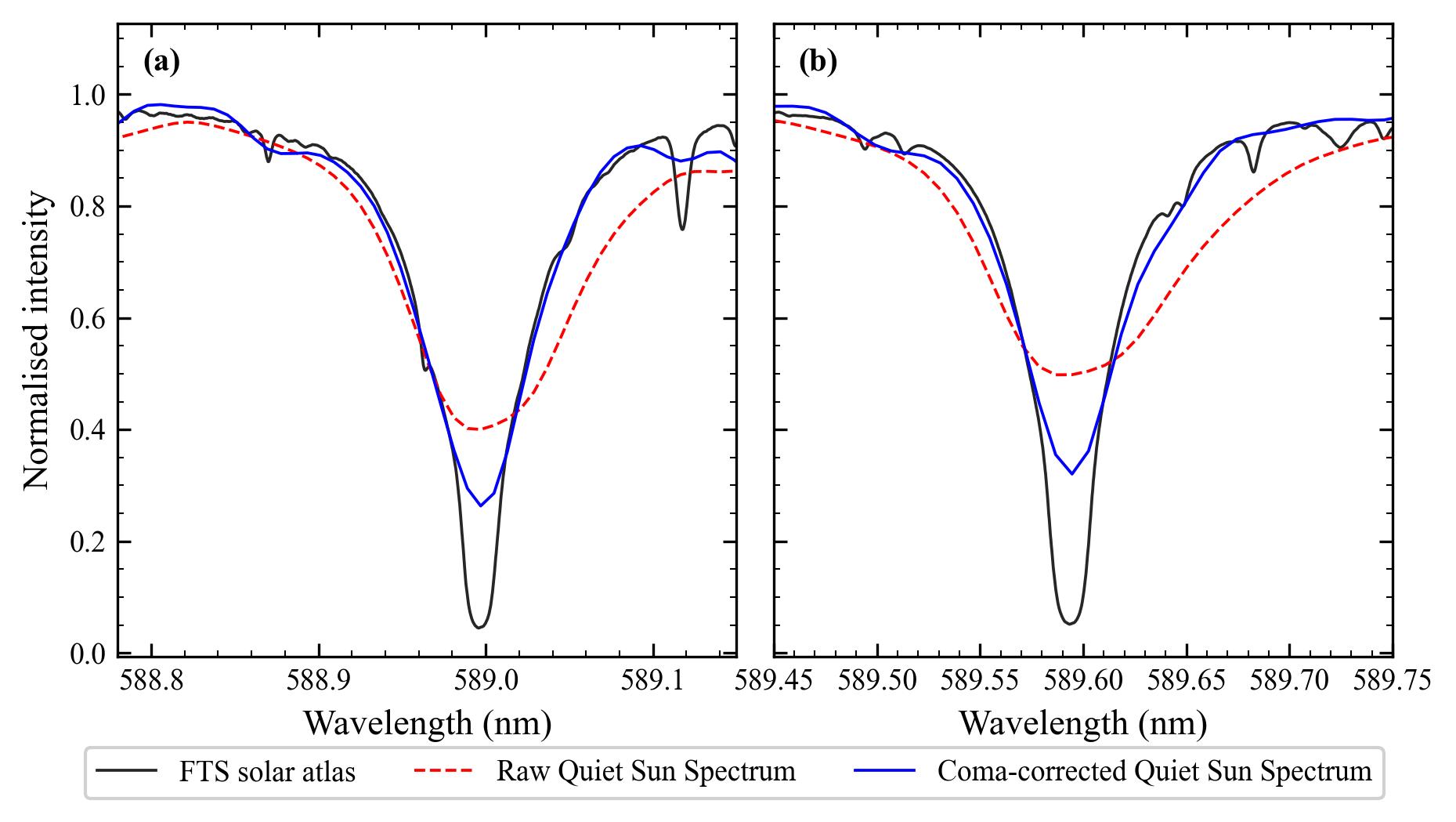}
	\centering
	\caption{A sample quiet-Sun spectrum observed with FRANCIS before (dashed red) and after (solid blue) application of the asymmetric line-spread-function deconvolution. Panel (a) shows the {\nadtwo} line and panel (b) shows the {\nadone} line. Both panels include the FTS solar atlas (solid black line) over the same wavelength range for comparison.
}
	\label{fig:spectra2}
\end{figure}

\subsection{Time series analysis}
\label{sec:timeseriesanalysis}
To investigate oscillatory signals within the sunspot atmosphere over the duration of the observations, both line-core Doppler and bisector velocity time series were computed for the {\nadone} and {\nadtwo} spectral lines. For line-core Doppler velocities, the wavelength of the turning point of the {\nadone} and {\nadtwo} lines were calculated using the WaLSA LineFit tool \citep{2026FrASS...Jafarzadeh_inprep}, as shown in Figure~{\ref{fig:spectra}}. This tool offers the ability to iteratively fit spectral lines using Voigt profiles and includes the ability to account for any line asymmetries that may exist within the absorption feature. The tool allows for tuning of the fitting window about the absorption feature. In this study, the fitting window was restricted to $\pm3$ pixels on either side of the reference wavelength to ensure the fit remained sensitive only to line core fluctuations. In addition to the average formation heights calculated from the contribution functions of the {\nadonetwo} lines (upper panels of Figure~{\ref{fig:contribution}}), further information can be extracted from the bisector wavelengths (i.e., at 75\% of the {\nadone} line depths). However, as noted in \citet{2023SoPh..298..146J}, the FRANCIS instrument has a slight increase in spectral broadening in the red wing of the observed spectral lines, which may be a consequence of an optical coma \citep{Tu2021} and/or a residual astigmatism caused by imperfect focusing of the tangential/sagittal focal planes provided by the toroidal mirror within the FRANCIS spectrograph \citep{Foreman:68, Lee:10}. Hence, before extraction of bisector metrics, it is important to first remove asymmetric instrumental broadening effects so they do not adversely affect the resulting bisector velocities computed. Therefore, a deconvolution kernel was produced to accurately re-map an example quiet Sun spectrum captured at disk center with the FRANCIS instrument onto the Fourier Transform Spectrometer \citep[FTS;][]{1978fsoo.conf...33B} solar atlas spectrum. This deconvolution kernel was subsequently applied to each of the science spectra, producing {\nadonetwo} spectra devoid of instrumental red wing asymmetries. 

\begin{figure}[!t]
	\includegraphics[width=1\textwidth]{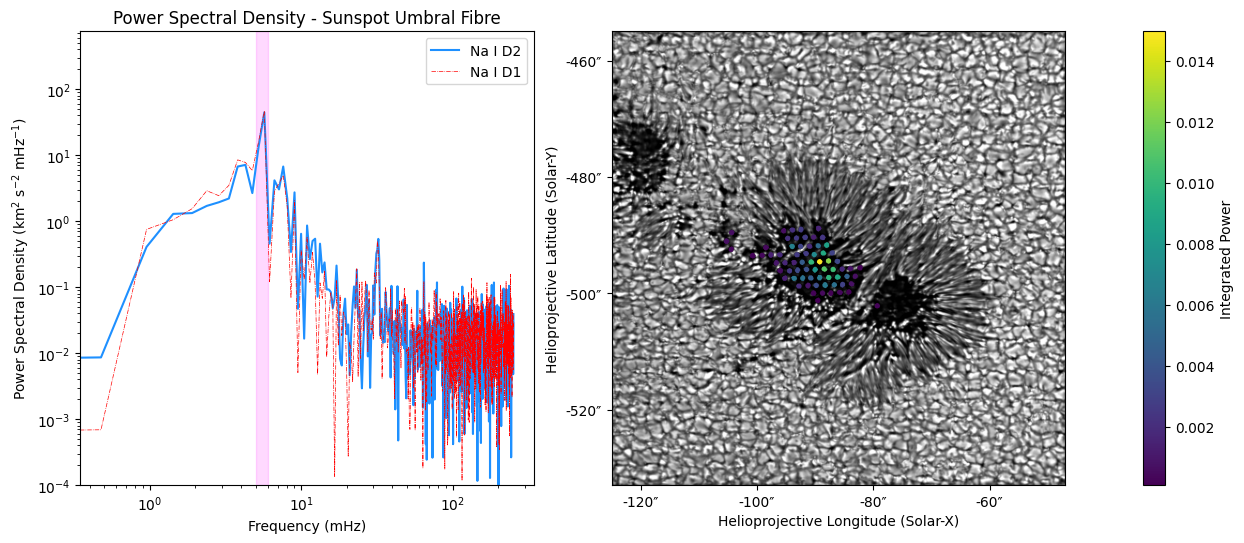}
	\centering
	\caption{Fourier power spectral densities (PSDs) of Doppler velocity time series for the {\nadone} (dashed orange line) and {\nadtwo} (solid blue line) spectral lines corresponding to a location at the center of the sunspot umbra. The vertical magenta band denotes the $5-6$~mHz frequency range where the power of both spectral diagnostics peak. The right panel shows the ROSA continuum context image, but with integrated {\nadtwo} spectral power from the FRANCIS umbral fibers between $5-6$~mHz overplotted using a purple-to-yellow color table. Hence, each circle corresponds to a unique FRANCIS fiber, while the color represents the $5-6$~mHz integrated power for that specific fiber location. Note that only fibers lying inside the sunspot umbra are included for visual clarity.}
	\label{fig:power}
\end{figure}

Comparisons of both the quiet-Sun spectrum from FRANCIS with the FTS solar atlas spectrum are presented in Figure~\ref{fig:spectra2}. The instrumental line spread function (LSF) was modeled as an asymmetric Gaussian kernel, parameterized by independent left- and right-hand standard deviations ($\sigma_\mathrm{left}$, $\sigma_\mathrm{right}$). The kernel parameters were determined empirically by iterative forward modeling: the FTS quiet-Sun solar atlas was convolved with candidate kernels across a grid of ($\sigma_\mathrm{left}$, $\sigma_\mathrm{right}$) values, and the result compared to the observed data. The resulting asymmetric kernel was then used to perform Wiener deconvolution in the Fourier domain, recovering a spectrum with the skew from the instrumental coma minimized to allow for subsequent bisector analysis. The resulting reduction in red-wing broadening can be seen in both panels of Figure~{\ref{fig:spectra2}}, where the corrected {\nadtwo} and {\nadone} profiles show improved agreement with the FTS solar atlas. We also note that spatially scattered photospheric light may contribute to the measured umbral spectra. The relatively high umbral continuum intensity, when normalized to the quiet-Sun continuum, suggests that a non-negligible scattered-light component cannot be ruled out. Such contamination would tend to dilute line depths and reduce the measured Doppler and intensity oscillation amplitudes, meaning that the derived wave powers and energy fluxes should be interpreted as lower-limit or order-of-magnitude estimates. However, moderate scattered-light contamination is not expected to generate the systematic spatial difference observed between near-zero phase behavior in the umbral core and frequency-dependent phase lags at the umbra-penumbra boundary. We therefore treat scattered light as an uncertainty on the absolute amplitudes and fluxes, rather than on the quantitative phase classification. Bisector wavelengths were subsequently calculated for the {\nadone} line at 75\% of its corresponding line depth to match the calculated formation heights depicted in the upper panels of Figure~{\ref{fig:contribution}}, i.e., 75\% of a quiescent {\nadone} line profile corresponds to the intensities found at the line core $-300$~m{\AA}, hence remaining consistent with the RH1.5D models and their computed formation heights. 

For each wavelength component ({\nadone} line core and 75\% of its line depth, plus the {\nadtwo} line core), and at each time step for the {\nadonetwo} spectral lines, the wavelengths were compared to their rest wavelengths, $\lambda_0$, of 588.994~nm and 589.592~nm to compute the relevant velocities as: 
\begin{equation*}
v_{\mathrm{LOS}} = c \, \frac{\lambda_{\mathrm{obs}} - \lambda_0}{\lambda_0} \ ,
\end{equation*} 
where $v_{\mathrm{LOS}}$ is the line-of-sight velocity calculated from the Doppler shift, $c$ is the speed of light, and $\lambda_\mathrm{obs}$ is the measured wavelength of the line core or bisector. Example velocity time series from the {\nadone} line core (i.e., 100\% of the corresponding line depth) and bisectors computed at 75\% of the {\nadone} line depth are shown in the lower-right panel of Figure~{\ref{fig:contribution}}.

\begin{figure}[t!]
	\includegraphics[width=0.8\textwidth]{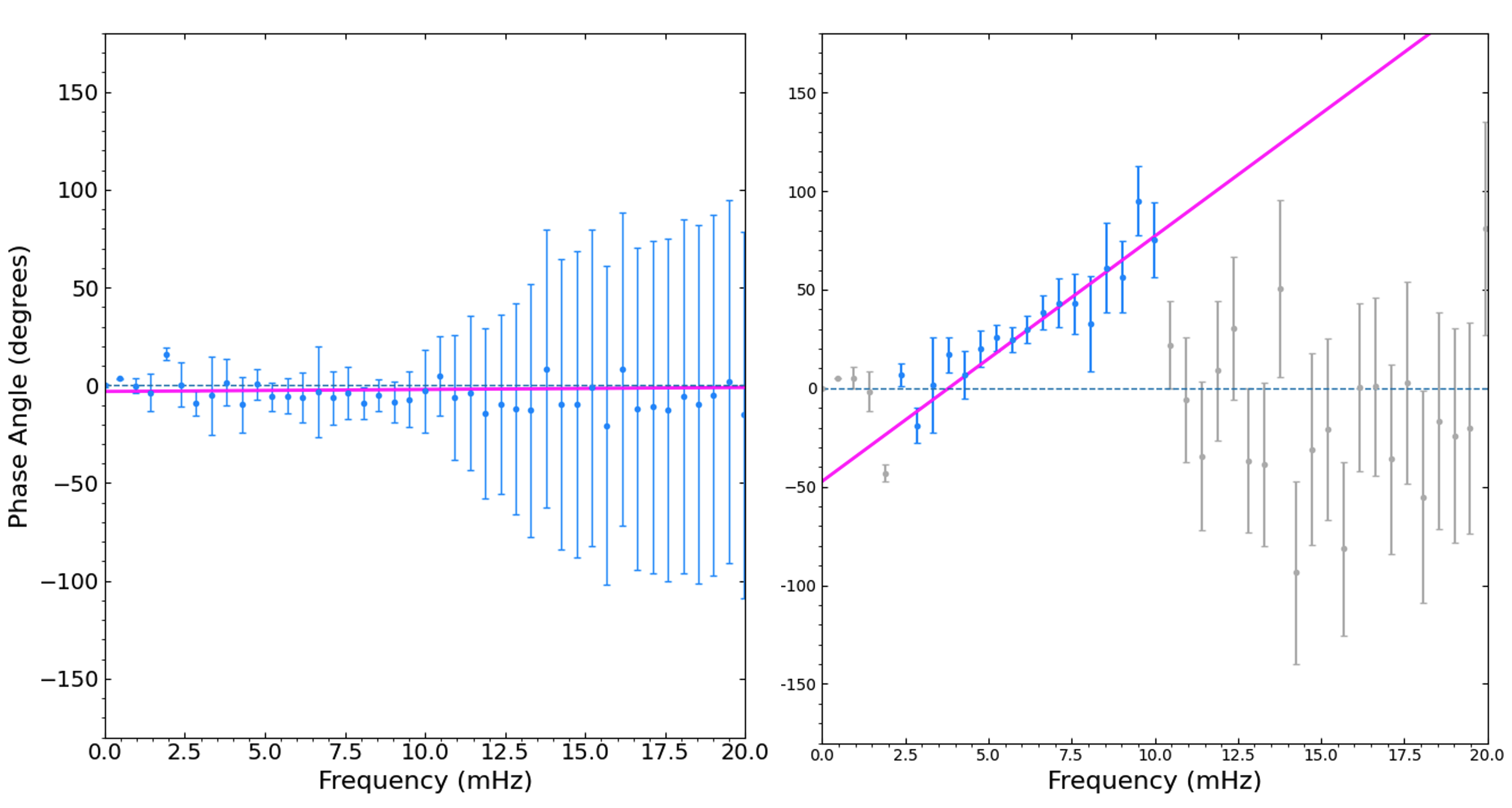}
	\centering
	\caption{Example phase spectra from wavelet cross correlation analysis between the {\nadone} and {\nadtwo} line core Doppler velocity time series. The left panel displays phase variations as a function of frequency for umbral core locations, while the right panel reveals phase variations as a function of frequency for FRANCIS fiber locations at the umbra-penumbra boundary. In each panel, the phase angle uncertainties correspond to the standard deviations of the phase angles determined by wavelet analysis, the horizontal dashed blue line represents $0^{\circ}$ phase, while the magenta line showcases a linear line of best fit applied to the measured phase values. The blue data points in the right panel represent phase angles that were utilised in the fitting of the gradient between phase angle and frequency, whereas the grey points are omitted.}
	\label{fig:phase_spec}
\end{figure}

To investigate the frequencies present within the Doppler velocity time series, the proven wave analysis techniques described in \citet{2023LRSP...20....1J} and \citet{2025NRvMP...5...21J} were employed. The time series were first detrended using a third-order polynomial to maximize stationarity, then a 90\% apodization filter was applied to mitigate edge effects arising due to the non-infinite duration of the time series. Next, power spectral densities (PSDs) were computed, with example PSDs for the {\nadone} and {\nadtwo} line core umbral velocity time series displayed in the left panel of Figure~{\ref{fig:power}}. Here, peak power is found at $\approx5.5$~mHz for both of the {\nadonetwo} lines, which remains consistent with previous {\nadonetwo} investigations \citep{1991A&A...241..212M, 2017ApJ...836...18C} and general chromospheric wave studies in sunspot umbrae \citep[e.g.,][to name but a few more recent examples]{2006ApJ...640.1153C, 2007PASJ...59S.631N, 2014A&A...561A..19Y, 2017ApJ...842...59J, 2018NatPh..14..480G, 2018ApJ...860...28H, 2020ApJ...892...49H, 2018ApJ...869..110S, 2020ApJ...896..150Y, 2021RSPTA.37900180S, 2022ApJ...927..201A, 2022ApJ...924..100C, 2023ApJ...944L..52C, 2023NatAs...7..856Y, 2025ApJ...985..256S}. A smaller power peak is also visible in the left panel of Figure~{\ref{fig:power}} at approximately $3-4$~mHz, which is more consistent with photospheric frequencies \citep{2008ApJ...676L..85K,  2018ApJ...852...15P, 2019ApJ...871..155R, 2026ApJ...997..232D}. However, as the dominant frequency for both the {\nadone} and {\nadtwo} line core velocity time series is $\approx5.5$~mHz, we are able to conclude that the {\nadonetwo} line cores more closely sample chromospheric plasma, which is supported by the line core contribution function shown in the upper-left panel of Figure~{\ref{fig:contribution}}. 

To examine the spatial distribution of the dominant $\approx5.5$~mHz wave power in the {\nadonetwo} line core velocity time series, we created power maps by integrating the PSD power between $5-6$~mHz (marked on the left panel of Figure~{\ref{fig:power}} using a vertical magenta band) and displaying this as a two-dimensional map placed on top of the ROSA continuum context image (right panel of Figure~{\ref{fig:power}}). The $\approx5.5$~mHz wave power that dominates the {\nadonetwo} line core velocity time series is concentrated towards the center of the umbra, gradually decreasing in magnitude towards the umbral/penumbral boundary. This is consistent with the work of \citet{2013A&A...554A.146K} and \citet{2024MNRAS.529..967S}, who were able to reveal that the largest wave power at $\approx5.5$~mHz in chromospheric diagnostics is concentrated towards the center of a sunspot umbra, where the magnetic field strengths are also at their strongest. However, we find a spatial offset of $\sim4000$~km ($\sim5.5$~arcseconds) between the location of peak wave power and the maximal umbral field strength, which according to the work of \citet{2021A&A...649A.169S} may indicate the presence of different MHD eigenmodes within the umbral atmosphere. However, to fully address this issue requires full Stokes spectropolarimetry to allow the magnetic oscillations to be benchmarked against their velocity counterparts, so this will form the basis of a future FRANCIS study once the polarization optics are fully operational. 

In the umbral PSDs displayed in the left panel of Figure~{\ref{fig:power}}, we note that an additional strong power peak is found at $\sim32$~mHz, which is around one order of magnitude stronger than the power demonstrated at neighboring frequencies. \citet{2020NatAs...4..220J} also found clear power peaks at higher ($\sim20$~mHz) frequencies within upper chromospheric He{\,}{\sc{i}}{\,}10830~{\AA} observations, which the authors speculated may be correlated with the value of the density scale height present within an umbral resonance cavity. Hence, if resonant waves are present within the sunspot atmosphere, these standing modes can be identified through phase relationships between neighbouring atmospheric layers, whereby propagating waves will have a frequency-dependent phase relationship, while standing modes will demonstrate in-phase oscillations across all frequencies \citep{2023LRSP...20....1J}. 

\begin{figure}[t!]
	\includegraphics[width=0.8\textwidth]{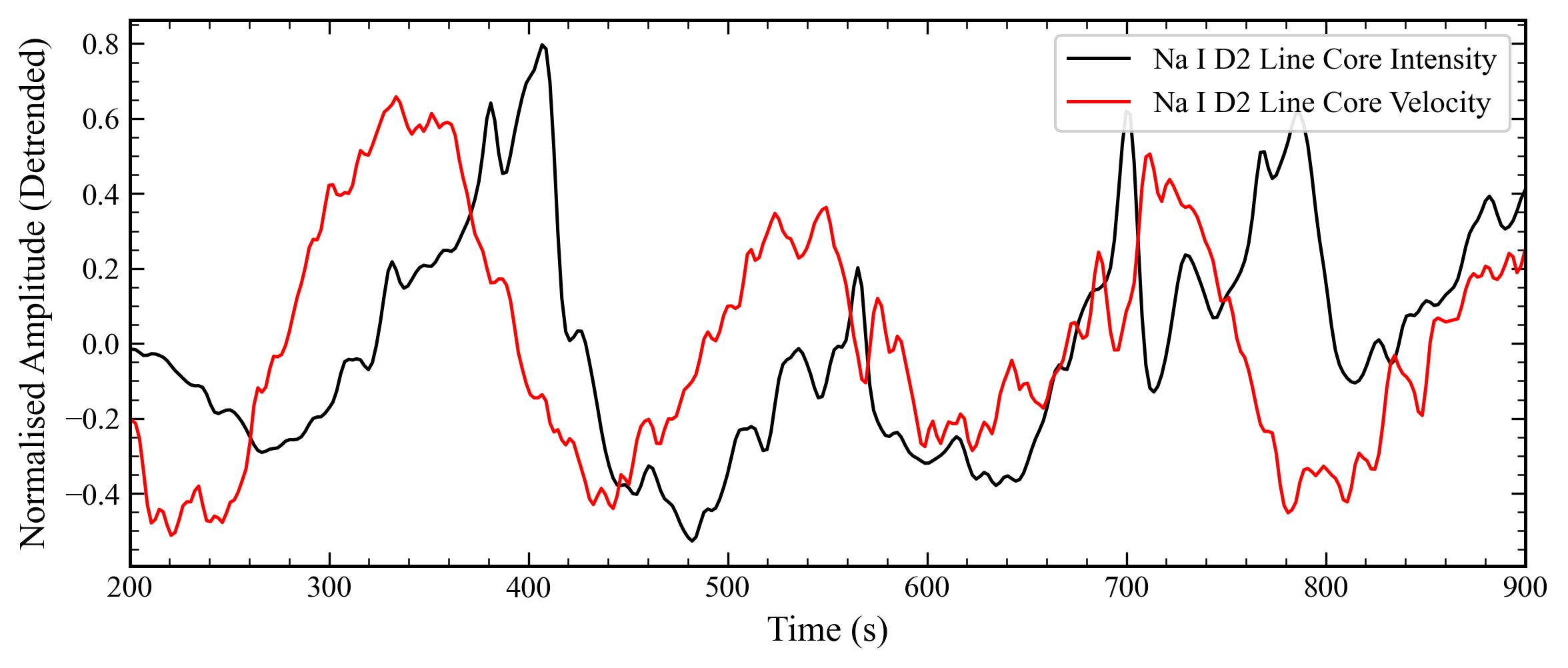}
	\centering
	\caption{Example time series for both {\nadtwo} line core Doppler velocity (red) and intensity (black). These time series were measured in a representative FRANCIS fiber sampling the umbral core, where the $5-6$~mHz oscillatory power is strongest.} 
	\label{fig:phase_v_t}
\end{figure}

In the present study, we have identified three unique wavelength positions (line core and 75\% of the line depth for the {\nadone} spectral line, alongside the line core of the {\nadtwo} spectral line) that, through examination of their respective contribution functions (see, e.g., the upper panels of Figure~{\ref{fig:contribution}}), provide us with three distinct formation heights over which we can examine the propagation characteristics of the embedded oscillations. To do so, we initially perform wavelet cross-correlation analysis \citep[following the methodology presented in][]{2023LRSP...20....1J, 2025NRvMP...5...21J} on the {\nadone} and {\nadtwo} line core velocity oscillations (coherence $> 0.8$), which have a height separation of $\sim 100$~km. The outputs of the wavelet cross correlation analysis provide us with a measurement of the phase lag between the two input time series, with wavelets preferred over conventional Fourier cross correlation analysis due to it providing increased number statistics across the full observational time frame \citep[see, e.g.,][]{2022ApJ...938..143G}. The resulting wavelet phase diagrams are subsequently condensed into one-dimensional phase plots (Figure~{\ref{fig:phase_spec}}) to remain consistent with previous examples of work in this area \citep[e.g.,][]{2006ApJ...640.1153C, 2017ApJ...847....5K}.

Phase spectra between the {\nadone} and {\nadtwo} line-core velocity time series are shown in Figure~{\ref{fig:phase_spec}}. The left panel represents {\nadonetwo} information extracted from 30 FRANCIS fibers closest to the center of the sunspot umbra, where the $5-6$~mHz wave power is strongest. The mean phase angle for these fibers remains close to $0^{\circ}$ over the frequency range where the phase uncertainties are smallest. Here, the error bars associated with each frequency correspond to the respective standard deviations in the computed (wavelet-derived) phase angles, following the approaches detailed in \citet{2022ApJ...938..143G}. Above 12.5~mHz, the uncertainties increase substantially, similar to the findings of \citet{2017ApJ...847....5K}, while near the dominant frequency of $\approx5.5$~mHz the uncertainties are as small as $\pm8^{\circ}$. {We interpret this near-zero phase behavior as consistent with standing-like oscillations in the upper-photospheric/lower-chromospheric layers sampled by the {\nadonetwo} line cores (i.e., $\sim750-850$~km). However, because the {\nadone} and {\nadtwo} contribution functions partially overlap and their nominal height separation is modest, the velocity-velocity phase relationship alone should not be regarded as unequivocal proof of a standing mode. Instead, it forms part of a broader set of evidence, together with the spatial concentration of $5-6$~mHz power in the umbral core and the contrast with propagating phase behavior at the umbra-penumbra boundary, which is in support of the increasing number of publications providing evidence for the existence of resonance cavities in sunspot umbral atmospheres \citep[e.g.,][]{2020NatAs...4..220J, 2021NatAs...5....5J, 2020ApJ...900L..29F, 2021A&A...645L..12F, 2025A&A...693A.165F, 2024MNRAS.529..967S}.}

On the other hand, the right panel of Figure~{\ref{fig:phase_spec}} represents {\nadonetwo} information extracted from 12~FRANCIS fibers located at the umbral/penumbral boundary. It is clear from this panel that the mean phase angle is a function of frequency, with phase angles $\approx0^{\circ}$ under 3~mHz \citep[where the waves are likely to be evanescent;][]{1975A&A....44..371D, 1979ApJ...231..570L, 1990A&A...236..509D, 2012ApJ...746..183J, 2016A&A...585A.110K}, then increasing monotonically up to phase angles of $\approx80^{\circ}$ at frequencies of 10~mHz. Similar to the umbral core phase spectra, frequencies beyond 12.5~mHz suffer from increased phase uncertainties, although within the frequency range of $1-10$~mHz the increase in phase angle with frequency is abundantly clear. This trend is representative of propagating waves \citep[see, e.g.,][]{2006ApJ...640.1153C, 2017ApJ...847....5K} manifesting towards the umbral/penumbral boundary, which may be related to the increasing magnetic field inclination angles in these locations providing the necessary environment for RPWs to propagate. 

While the dominant $\approx5.5$~mHz power identified in the {\nadonetwo} line-core velocity time series is consistent with chromospheric wave activity in sunspot umbrae, we also consider the potential influence of umbral flashes (UFs). Early work noted a correlation between oscillations in the Ca~{\textsc{ii}}~K line core and {\nadonetwo} measurements \citep{1981A&A...102..147K}, motivating us to examine whether UF activity is present in our observations. We find no clear evidence of UFs in the present data. The {\nadonetwo} velocity time series exhibit smooth oscillatory signatures, with no indication of sawtooth temporal profiles or asymmetric line-core emission reversals characteristic of chromospheric shock-driven UF events \citep{1969SoPh....7..351B}. This is further supported by the formation properties of the {\nadonetwo} lines, whose line-core contribution functions peak at geometric heights $\lesssim$1000~km, spanning the upper photosphere and base of the chromosphere. These heights are below the atmospheric layers in which UF thermal perturbations are typically most pronounced \citep{2018NatPh..14..480G}. We therefore conclude that umbral flashes are unlikely to significantly influence the {\nadonetwo} line-core velocity measurements presented here, although we cannot exclude UF signatures in higher-forming diagnostics.

{Following the methods of \citet{2020ApJ...900L..29F}, we also performed velocity-intensity (V--I) phase analysis on the {\nadtwo} line-core time series, using intensity fluctuations as a proxy for temperature perturbations. Across the full observing sequence, the phase spectra do not show a persistent frequency-independent $\pi/2$ phase difference, nor the corresponding jump to $-\pi/2$~radians associated with resonance nodes. However, during the coherent interval between approximately $200-900$~s shown in Figure~{\ref{fig:phase_v_t}}, the Doppler velocity leads the intensity by approximately $\pi/2$~radians. This behavior is consistent with the velocity-temperature phase relationship expected for a chromospheric resonance cavity, but because it is not persistent across the full observing sequence, we treat it as supporting evidence rather than a decisive diagnostic of resonant behavior.}

To better examine the propagation characteristics of the waves found towards the umbral/penumbral boundary, we also employ bisector analysis corresponding to 75\% of the {\nadone} line depth to provide us with an additional formation height to better track the wave propagation. As documented in the upper panels of Figure~{\ref{fig:contribution}}, this bisector wavelength position provides an estimated formation height of $\approx355$~km, which can then be combined with the $\approx750$~km and $\approx850$~km formation heights of the {\nadone} and {\nadtwo} line cores, respectively, to provide 3 distinct layers from which we can probe the embedded wave behavior using our previously defined cross correlation analyses. We note, however, that these geometric heights are derived from an umbral model atmosphere and therefore provide only first-order reference values at the umbra--penumbra boundary; consequently, the phase speeds, energy fluxes, and damping lengths derived below should be interpreted as approximate, model-dependent estimates. Across these formation heights, the phase speed of the wave (estimated by the sound speed, $c_s$, due to the magnetoacoustic nature of the waves), can be calculated from the derived phase angles and estimated height separations following \citep{2023LRSP...20....1J}, 
\begin{equation*}
{c_s} = f\,\Delta z \times \frac{360}{\Delta \phi} \ ,
\end{equation*} 
where $\Delta \phi$ denotes the phase difference (in degrees), $f$ is the oscillation frequency, and $\Delta z$ is the height separation between the two formation layers. The local sound speed, $c_s$, can hence be obtained from the gradient of the phase/frequency relationship, as seen in Figure~{\ref{fig:phase_spec}}. From the example phase spectra shown in the right panel of Figure~{\ref{fig:phase_spec}}, corresponding to the phase lag between the {\nadonetwo} line core velocity time series (corresponding to formation heights of 750~km and 850~km, respectively), we calculate a sound speed value of $c_s = 9.3 \pm 1.2$~km{\,}s$^{-1}$. For the phase spectra between 75\% of the {\nadone} line depth and the line core (corresponding to formation heights of 355~km and 750~km, respectively), we find a slightly higher sound speed of $c_s = 11.8 \pm 2.0$~km{\,}s$^{-1}$.

To verify these values, we also compute the sound speed using temperature values extracted from the Maltby `M' sunspot umbral core model \citep{1986ApJ...306..284M} at the three representative formation heights employed here: 355~km, 750~km, and 850~km. The theoretical sound speed was calculated according to the methods detailed in \citet{2017ApJ...847....5K} as,
\begin{equation*}
c_s = \sqrt{\frac{\gamma R T}{\mu}} \ ,
\end{equation*}
where $\gamma$ is the polytropic index, $R$ is the gas constant, and $\mu$ is the mean molecular weight. In this study we consider $\gamma = 5/3$, $R = 8.314 \times 10^{7}$~erg{\,}K$^{-1}${\,}mol$^{-1}$, and $\mu = 0.61$ \citep{1993Sci...261..239M}. With these values, we compute sound speeds as 8.8~km{\,}s$^{-1}$, 9.1~km{\,}s$^{-1}$, and 9.7~km{\,}s$^{-1}$ at atmospheric heights of 355~km, 750~km, and 850~km, respectively. These sound speeds are in broad agreement with the observationally derived values of $c_s = 11.8 \pm 2.0$~km{\,}s$^{-1}$ (atmospheric heights of $355\rightarrow750$~km) and $c_s = 9.3 \pm 1.2$~km{\,}s$^{-1}$ (atmospheric heights of $750\rightarrow850$~km). The slightly higher sound speed calculated between the {\nadone} line wing and {\nadone} line core is likely the result of more uncertainty associated with the formation height linked to 75\% of the {\nadone} line depth, which translates into increased velocity uncertainty in the estimated sound speed between these atmospheric layers. Nevertheless, a sound speed of $\sim9.6$~km{\,}s$^{-1}$ can be used for subsequent energy flux calculations due to its consistent magnitude between both model atmosphere calculations and observational measurements. 

With the wave propagation speed calculated, it is possible to estimate the energy flux of the propagating waves seen at the umbra/penumbra boundary across the three atmospheric heights under investigation following \citet{2021RSPTA.37900172G} as,  
\begin{equation*}
E = \rho \langle \delta v^2 \rangle c_s \ ,
\end{equation*} 
where $E$ is the wave energy flux, $\rho$ is the plasma mass density, and $\langle \delta v^2 \rangle$ is the mean square velocity fluctuation identified from the input velocity time series. The densities obtained from the Maltby `M' umbral core model at $355$~km, $750$~km, and $850$~km are $2.928 \times 10^{-5}$~kg{\,}m$^{-3}$, $6.731 \times 10^{-7}$~kg{\,}m$^{-3}$, and $2.579 \times 10^{-7}$~kg{\,}m$^{-3}$, with root mean square velocity amplitudes in the corresponding umbral time series measured to be $0.309$~km{\,}s$^{-1}$, $0.783$~km{\,}s$^{-1}$, and $1.037$~km{\,}s$^{-1}$, respectively. In the lower atmospheric layer spanning $355~{\mathrm{km}}\rightarrow750~{\mathrm{km}}$, this provides an energy flux of $1.267 \times 10^{4}$~W{\,}m$^{-2}$. At the higher atmospheric heights of between $750~{\mathrm{km}}\rightarrow850~{\mathrm{km}}$, we find the energy flux decreases to $3.145 \times 10^{3}$~W{\,}m$^{-2}$, where this approximate factor-of-four decrease is consistent with previous studies of wave propagation in magnetic structures \citep{2017ApJ...847....5K, 2021RSPTA.37900172G}. 

With the energy flux estimated for two distinct atmospheric layers, the damping length equation provided in \citet{2019FrASS...6...57S} can be adapted to its energy flux equivalent to give,
\begin{equation*}
E(h) = E_0 \, e^{-2h/L_d} \ ,
\end{equation*}
where $E(h)$ is the energy flux at height $h$, $E_{0}$ is the reference energy flux, and $L_d$ is the damping length. The energy flux gradient, as a function of atmospheric height, therefore provides an estimate of the damping length, which corresponds to $L_d \approx 363$~km in our present study. Our damping length is slightly higher than the damping lengths previously calculated for both photospheric quiet Sun and magnetic pores \citep[$L_d \approx 210$~km and $L_d \approx 268$~km, respectively;][]{2021RSPTA.37900172G}. However, as our present study focuses on atmospheric heights synonymous with the upper photosphere and lower chromosphere, it is expected that the damping lengths will become larger as a result of the comparable increase in the density scale height \citep{2014ApJ...789..118K, 2019FrASS...6...57S}.

The coexistence of standing-like and propagating phase behavior also helps reconcile the present results with earlier {\nadonetwo} studies that interpreted phase differences as evidence for upward propagation. In the present observations, upward propagation is indeed detected toward the umbra-penumbra boundary, where the magnetic field is more inclined and lower-frequency waves may propagate more readily. The near-zero phase behavior is instead concentrated in the umbral core, where the strongest $5-6$~mHz power is observed and where the magnetic field is close to vertical. Thus, the FRANCIS IFU observations do not imply that the {\nadonetwo} atmosphere is globally standing. Instead, they show that propagating and standing-like signatures coexist spatially within the same sunspot atmosphere.

\section{Concluding Remarks}
In this study we have exploited the new integral field unit, FRANCIS, to explore the oscillatory behavior of the {\nadonetwo} spectral lines simultaneously across a two-dimensional sunspot umbral field of view. With simultaneous observations of both spectral lines, alongside supporting bisector velocity analysis, we have demonstrated a mixture of standing and propagating waves in the atmosphere above the sunspot. 

First, at the umbra-penumbra boundary, we find evidence of propagating waves with phase speeds of $11.8 \pm 2.0$~km{\,}s$^{-1}$ between heights of 355~km and 750~km, and $9.3 \pm 1.2$~km{\,}s$^{-1}$ between heights of 750~km and 850~km. Due to these waves propagating along magnetic field lines with larger inclination angles compared to those at the center of the umbra, we propose that these may indicate the lower atmospheric components of RPWs \citep{2013ApJ...779..168J}. By combining these inferred phase speeds with mean square velocity amplitudes at each height, we determine energy fluxes on the order of $1.267 \times 10^{4}$~W{\,}m$^{-2}$ and $3.145 \times 10^{3}$~W{\,}m$^{-2}$ at the lower and higher atmospheric layers, respectively. Such a decrease in energy flux as the waves propagate upward suggests a damping length of $L_{d} \approx 363$~km, which is marginally larger than the damping lengths computed in photospheric wave observations \citep{2021RSPTA.37900172G}, although still consistent with the overall density scale height expected in the upper photosphere and lower chromosphere. 

Second, and perhaps most importantly, we utilize the simultaneous imaging and spectroscopy capabilities of an integral field unit to demonstrate that chromospheric oscillations with a peak frequency of $\approx5.5$~mHz dominate the center of the sunspot umbra. Unlike the propagating waves at the umbra-penumbra boundary, these central umbral oscillations exhibit near-zero velocity phase differences over the frequency range where the phase uncertainties are smallest. This behavior is consistent with standing-like oscillations at the lower-chromospheric formation heights sampled by the {\nadonetwo} spectral lines, supporting the interpretation of a wave-amplifying resonance cavity. The PSDs of these locations also give rise to a secondary peak at $\approx32$~mHz, which may be correlated with the density scale height within an umbral resonance cavity \citep{2020NatAs...4..220J, 2021NatAs...5....5J}.

These initial observations using FRANCIS demonstrate the capability of a fiber-fed IFU to simultaneously resolve spatial and spectral information at high cadence, revealing vertical atmospheric wave structuring and the coexistence of propagating and standing-like slow magnetoacoustic signatures. Future work will employ the FRANCIS polarization optics to search for magnetic wave signatures in complete Stokes~$I/Q/U/V$ spectra, and will use modern machine-learning and artificial-intelligence approaches to perform rapid spectropolarimetric inversions \citep[e.g.,][]{2025ApJ...988....9C, 2026ApJ...996...63C}. The analysis techniques applied here are not exclusive to FRANCIS, but are directly relevant to future studies with other solar IFUs. In particular, higher-spatial-resolution observations from facilities such as DKIST, including DL-NIRSP integral-field spectroscopy \citep{dlnirsp}, will provide an important route toward resolving wave behavior across a broader range of atmospheric heights.

\section*{Conflict of Interest Statement}
The authors declare that the research was conducted in the absence of any commercial or financial relationships that could be construed as a potential conflict of interest.

\section*{Author Contributions}
GC: Data curation, Formal analysis, Investigation, Methodology, Software, Validation, Visualization, Writing -- original draft;
DBJ: Conceptualization, Data curation, Funding acquisition, Project administration, Resources, Supervision, Writing -- original draft;
SJ: Investigation, Methodology, Software, Validation, Visualization, Writing -- original draft;
MB: Formal analysis, Investigation, Methodology, Writing -- original draft;
SDTG: Conceptualization, Data curation, Investigation, Supervision, Writing -- original draft;
MS: Resources, Software, Writing -- original draft;
HNS: Resources, Software, Writing -- original draft;
DJC: Investigation, Resources, Writing -- original draft;
LEAV: Methodology, Resources, Supervision, Writing -- original draft;
ADL: Resources, Supervision, Writing -- original draft;
FLG: Supervision, Writing -- original draft.

\section*{Funding}
UK Science and Technology Facilities Council (STFC) consolidated grants ST/T00021X/1 and ST/X000923/1.
UK STFC PATT Travel Grant UKRI372.
Leverhulme Trust Research Project Grant RPG-2019-371. 
UK Space Agency National Space Technology Programme grant SSc-009.
The Royal Society award no. Hooke18b/SCTM.
NASA grants 19-HSODS-004 and 21-SMDSS21-0047.
International Space Science Institute (ISSI) Team 502.

\section*{Acknowledgments}
GC, DBJ and SJ wish to thank the UK Science and Technology Facilities Council (STFC) for the consolidated grants ST/T00021X/1 and ST/X000923/1, alongside the PATT Travel Grant UKRI372.
DBJ acknowledges support from the Leverhulme Trust via the Research Project Grant RPG-2019-371. 
DBJ and SDTG also acknowledge funding from the UK Space Agency via the National Space Technology Programme (grant SSc-009).
SJ acknowledges support from the Rosseland Centre for Solar Physics, University of Oslo, Norway, and the Max Planck Institute for Solar System Research, Germany.
MB acknowledges that this publication was produced while attending the PhD programme in Space Science and Technology at the University of Trento, Cycle XXXIX, with the support of a scholarship financed by the Ministerial Decree no. 118 of 2nd March 2023, based on the NRRP - funded by the European Union - NextGenerationEU - Mission 4 ``Education and Research'', Component 1 ``Enhancement of the offer of educational services: from nurseries to universities'' - Investment 4.1 ``Extension of the number of research doctorates and innovative doctorates for public administration and cultural heritage'' - CUP E66E23000110001.
DJC acknowledges partial support of this project from NASA grants 19-HSODS-004 and 21-SMDSS21-0047.
We wish to acknowledge scientific discussions with the Waves in the Lower Solar Atmosphere (WaLSA; \href{https://WaLSA.team}{www.WaLSA.team}) team, which has been supported by the Research Council of Norway (project no. 262622), The Royal Society \citep[award no. Hooke18b/SCTM;][]{2021RSPTA.37900169J}, and the International Space Science Institute (ISSI) in Bern through ISSI International Team project 502 ``WaLSA: Waves in the Lower Solar Atmosphere at High Resolution''.


\section*{Data Availability Statement}
The data used in this paper are from the observing campaign entitled {\textit{`FRANCIS: commissioning a next generation IFU on the Dunn Solar Telescope'}} (Principal Investigator: D.B. Jess), which employed the ground-based Dunn Solar Telescope, USA, during August 2022. Additional supporting observations were obtained from the publicly available NASA Solar Dynamics Observatory (\href{https://sdo.gsfc.nasa.gov}{https://sdo.gsfc.nasa.gov}) data archive, which can be accessed via \href{http://jsoc.stanford.edu/ajax/lookdata.html}{http://jsoc.stanford.edu/ajax/lookdata.html}. The ground-based data obtained during 29 August 2022 are several TB in size and cannot be hosted on a public server. However, all data supporting the findings of this study are available directly from the authors on request.

\bibliographystyle{Frontiers-Harvard} 
\bibliography{main}

\end{document}